\documentstyle[12pt]{article}
\begin{document}

\title{ON THE INFLUENCE OF GRAVITATIONAL RADIATION ON A GYROSCOPE }

\author{L. Herrera\thanks{Also at UCV, Caracas, Venezuela. E-mail address:
lherrera@gugu.usal.es}\\Area de F\'\i
sica Te\'orica. Facultad de Ciencias.\\ Universidad de Salamanca. 37008
Salamanca, Espa\~na. \\ and\and J. L. Hern\'andez
Pastora \\Area de F\'\i sica Te\'orica. Facultad de Ciencias.\\ Universidad de
Salamanca. 37008 Salamanca, Espa\~na.}

\date{}
\maketitle

\begin{abstract}
We calculate the precession of a gyroscope at rest in a Bondi spacetime.
It is shown that, far from the source, the leading term in the
 rate of precession of the gyroscope is simply expressed through
the news function of the system, and vanishes if and only if there is no news.
Rough estimates are presented, illustrating the order of magnitude of the
expected effect for different scenarios.
It is also shown from the next  order term ( ${\displaystyle \frac 1{r^2}}$)
that non-radiative (but
time dependent) spacetimes will produce a gyroscope precession of that order,
providing thereby ``observational'' evidence for the violation of the Huygens's
principle.
\end{abstract}

\newpage
\section{Introduction}
The theoretical description and the experimental observation of
gravitational radiation are among the most relevant challenges
confronting general relativity.

A great deal of work has been done so far in order to provide
a consistent framework for the study of such phenomenon. Also,since Weber's
pioneering work \cite{weber}
important collaboration efforts have been carried on, and are
now under consideration, to put in evidence gravitational waves (see
\cite{michelson}, \cite{smarr},
\cite{brillet},\cite{will}, and references therein).

It is the purpose of this work to evaluate the influence of gravitational
radiation on a gyroscope.This idea is not new,
 in fact some years ago, Chaboyer and Henriksen \cite{chaboyer}
put forward the possibility of detecting gravitational radiation by means
of an orbital laser gyroscope. In such experiment
the presence of gravitational radiation is brought out by the differential
effect that radiation has on the paths
of the photons in the rotating frame of reference.

In this work we shall calculate  the rate of precession of a gyroscope in
the field of gravitational radiation.
To do so we shall use the Bondi's formalism \cite{Boal}
which has, among other things, the virtue of providing a clear and precise
criterion for the existence of gravitational radiation (see also \cite{JaNe}).
Namely, if the news function is zero over a time interval,
then there is no radiation during that interval. Also, the present approach
has the advantage of providing
a very simple expression linking an``observable'' (at least in principle)
quantity (the rate of precession of
a gyroscope) with the emission rate of gravitational radiation.

The formalism has as its main drawback \cite{BoNP} the fact that
it is based on a series expansion which could not give closed
solutions and which raises unanswered questions about convergence
and appropriateness of the expansion.

However since we shall assume the gyroscope to be very far
from the source, we shall use in our calculations only the
leading terms in the expansion of metric functions. Furthermore,
since the source is assumed to radiate during a finite interval,
then no problem of convergence appears \cite{Bo}.

We shall see that the leading term of the rate of precession of
the gyroscope ($\Omega$) is expressed through the news function in such a
way that it will vanish if and only if there is no news (no radiation).

In the special case of the quadrupole radiation (in the linear approximation),
the rate of precession may be expressed through the third time
derivative of the quadrupole moment, or alternatively, through
the rate of loss of the mass function.

 Next we present the  order (${\displaystyle \frac 1{r^2}}$) of
$\Omega$. As
we shall see, it contains terms with news, together with a time
dependent term not involving news. This last term represents the class of
non-radiative motions discussed by Bondi \cite{Boal} and may be thought to
correspond to the tail of the wave, appearing after the radiation process
\cite{BoNP}. The obtained expression allows for ``measuring'' (in a
gedanken experiment, at least) the
wave-tail field. This in turn implies that observing the gyroscope, for a
period of time from an initial static situation until after the vanishing of
the news, should allow for an unambiguous identification of a gravitational
radiation process.

In the next section we briefly present the Bondi's formalism.
The expression for the rate of precession of the gyroscope in
the Bondi metric is calculated in sections 3 and 4, and estimates for the
rate of
precession in different scenarios are presented in section 5. Finally the
results
are discussed in the last section.
\section{The Bondi's Formalism}
The general form of an axially symmetric asymptotically flat
metric given by Bondi is \cite{Boal}
\begin{eqnarray}
ds^2 & = & \left(\frac{V}{r} e^{2\beta} - U^2 r^2 e^{2\gamma}\right) du^2
+ 2 e^{2\beta} du dr \nonumber \\
& + & 2 U r^2 e^{2\gamma} du d\theta
- r^2 \left(e^{2 \gamma} d\theta^2 + e^{-2\gamma} \sin^2{\theta} d\phi^2\right)
\label{Bm}
\end{eqnarray}
where $V, \beta, U$ and $\gamma$ are functions of
$u, r$ and $\theta$.

We number the coordinates $x^{0,1,2,3} = u, r, \theta, \phi$ respectively.
$u$ is a timelike coordinate such that $u=constant$ defines a null surface.
In flat spacetime this surface coincides with the null light cone
open to the future. $r$ is a null coordinate ($g_{rr}=0$) and $\theta$ and
$\phi$ are two angle coordinates (see \cite{Boal} for details).

Regularity conditions in the neighborhood of the polar axis
($\sin{\theta}=0$), implies that
as $\sin{\theta}->0$
\begin{equation}
V, \beta, U/\sin{\theta}, \gamma/\sin^2{\theta}
\label{regularity}
\end{equation}
each equals a function of $\cos{\theta}$ regular on the polar axis.

The four metric functions are assumed to be expanded in series of $1/r$,
then using field equations Bondi gets

\begin{equation}
\gamma = c(u,\theta) r^{-1} + \left(C(u,\theta) - \frac{1}{6} c^3\right) r^{-3}
+ ...
\label{ga}
\end{equation}
\begin{equation}
U = - \left(c_\theta + 2 c \cot{\theta}\right) r^{-2} + ...
\label{U}
\end{equation}
\begin{eqnarray}
V & = & r - 2 M(u,\theta)\nonumber \\
& - & \left( N_\theta + N \cot{\theta} -
c_{\theta}^{2} - 4 c c_{\theta} \cot{\theta} -
\frac{1}{2} c^2 (1 + 8 \cot^2{\theta})\right) r^{-1} + ...
\label{V}
\end{eqnarray}
\begin{equation}
\beta = - \frac{1}{4} c^2 r^{-2} + ...
\label{be}
\end{equation}
where letters as subscripts denote derivatives, and

\begin{equation}
4C_u = 2 c^2 c_u + 2 c M + N \cot{\theta} - N_\theta
\label{C}
\end{equation}

The three functions $c, M$ and $N$ are further
related by the supplementary conditions
\begin{equation}
M_u = - c_u^2 + \frac{1}{2}
\left(c_{\theta\theta} + 3 c_{\theta} \cot{\theta} - 2 c\right)_u
\label{M}
\end{equation}
\begin{equation}
- 3 N_u = M_\theta + 3 c c_{u\theta} + 4 c c_u \cot{\theta} + c_u c_\theta
\label{N}
\end{equation}

In the static case $M$ equals the mass of the system whereas $N$ and $C$
are closely related to the dipole and quadrupole moment respectively.

Next, Bondi defines the mass $m(u)$ of the system as
\begin{equation}
m(u) = \frac{1}{2} \int_0^\pi{M(u,\theta) \sin{\theta} d\theta}
\label{m}
\end{equation}
which by virtue of (\ref{M}) and (\ref{regularity}) yields
\begin{equation}
m_u = - \frac{1}{2} \int_0^\pi{c_u^2 \sin{\theta} d\theta}
\label{muI}
\end{equation}

Let us now recall the main conclusions emerging from the Bondi's approach.
\begin{enumerate}
\item If $\gamma, M$ and $N$ are known for some $u=a$(constant) and
$c_u$ (the news function) is known for all $u$ in the interval
$a \leq u \leq b$,
then the system is fully determined in that interval. In other words,
whatever happens at the source, leading to changes in the field,
it can only do so by affecting $c_u$ and viceversa. At the
light of this comment the relationship between news function
and the occurrence of radiation becomes clear.
\item As it follows from (\ref{muI}), the mass of a system is constant
if and only if there are no news.
\end{enumerate}

In the next section we calculate the rate of precession of a
gyroscope at rest in the frame of (\ref{Bm})

\section{The gyroscopic precession}
Let us start by defining the vorticity vector, which as usual
is given by (in relativistic units)
\begin{eqnarray}
\omega^\alpha & = & \frac{1}{2\sqrt{-g}} \epsilon^{\alpha\beta\gamma\delta}
{u}_{\beta} \omega_{\gamma\delta} \nonumber \\
& = & \frac{1}{2\sqrt{-g}} \epsilon^{\alpha\beta\gamma\delta}
{u}_{\beta} {u}_{\gamma,\delta}
\label{vv}
\end{eqnarray}
where the vorticity tensor is given by
\begin{equation}
\omega_{\alpha\beta} = u_{[\alpha;\beta]} - \dot{u}_{[\alpha}u_{\beta]}
\label{vt}
\end{equation}
and $u_\beta$ denotes the four-velocity vector.

Now, for an observer at rest in the frame of (\ref{Bm}), the four-velocity
vector has components
\begin{equation}
u_\alpha = \left(A, \frac{e^{2\beta}}{A}, \frac{U r^2 e^{2\gamma}}{A}, 0\right)
\label{fv}
\end{equation}
with
\begin{equation}
A \equiv \left(\frac{V}{r} e^{2\beta} - U^2 r^2 e^{2\gamma}\right)^{1/2}
\label{A}
\end{equation}
using (\ref{fv}) and
\begin{equation}
\sqrt{-g} = e^{2\beta} r^2 \sin{\theta}
\label{det}
\end{equation}
in (\ref{vv}), we easily obtain
\begin{equation}
\omega^\alpha = \left(0, 0, 0, \omega^3\right)
\label{oma}
\end{equation}
with
\begin{eqnarray}
\omega^3 & = & -\frac{e^{-2\beta}}{2 r^2 \sin{\theta}}
\{
2 \beta_\theta e^{2\beta}
- \frac{2 e^{2\beta} A_\theta}{A}
- \left(U r^2 e^{2\gamma}\right)_r \nonumber \\
& + & \frac{2 U r^2 e^{2\gamma}}{A} A_r
+ \frac{e^{2\beta} \left(U r^2 e^{2\gamma}\right)_u}{A^2}
- \frac{U r^2 e^{2\gamma}}{A^2} 2 \beta_u e^{2\beta}
\}
\label{om3}
\end{eqnarray}
and for the absolute value of $\omega^\alpha$ we get
\begin{eqnarray}
\Omega & \equiv & \left(- \omega_\alpha \omega^\alpha\right)^{1/2} =
 \frac{e^{-2\beta -\gamma}}{2 r}
\{2 \beta_\theta e^{2\beta} - 2 e^{2\beta} \frac{A_\theta}{A}
 -  \left(U r^2 e^{2\gamma}\right)_r
\nonumber \\
& + & 2 U r^2 e^{2\gamma} \frac{A_r}{A}
 +  \frac{e^{2\beta}}{A^2} \left(U r^2 e^{2\gamma}\right)_u
 - 2 \beta_u \frac{e^{2\beta}}{A^2} U r^2 e^{2\gamma}
\}
\label{OM}
\end{eqnarray}
Feeding back (\ref{ga})--(\ref{be}) into (\ref{OM}) and
keeping only the leading term, we obtain
\begin{equation}
\Omega = - \frac{1}{2r} \left(c_{u \theta} + 2 c_u \cot{\theta}\right)
+ O(r^{-n}) \; ; \; n>1
\label{Om}
\end{equation}

Now, since $\Omega$ measures the rate of rotation with respect
to proper time of world lines of points at rest in the frame
of (\ref{Bm}), relative to the local compass of inertia, then
$-\Omega$ describes the rotation of the compass of inertia (`` the
gyroscope'') with respect to reference particles at rest in the
frame of (\ref{Bm}) (see \cite{RiPe} for detailed discussion on this point).

Therefore, up to order $1/r$, the gyroscope will precess as long as
the system radiates ($c_{u} \not= 0$). Observe that if
\begin{equation}
c_{u\theta} + 2 c_u \cot{\theta} = 0
\label{if}
\end{equation}
then
\begin{equation}
c_u = \frac{F(u)}{\sin^2{\theta}}
\label{cu}
\end{equation}
which implies
\begin{equation}
F(u) = 0  \Longrightarrow c_u = 0
\label{F}
\end{equation}
in order to insure regularity conditions, mentioned above,
 in the neighbourhood
of the polar axis ($\sin{\theta} = 0$) .
Thus the leading term in (\ref{Om}) will vanish if and only if $c_u = 0$.

If the system radiates during an interval of time $\Delta u$, then
the change of orientation of the gyroscope, for that period, is given by
\begin{equation}
\Delta\phi = - \Omega \Delta u \left(
\frac{V}{r} e^{2\beta} - U^2 r^2 e^{2\gamma}\right)^{1/2}
\label{Dfi}
\end{equation}
or, up to terms of order $1/r$
\begin{equation}
\Delta \phi \approx \frac{1}{2r} \left(c_{u\theta} +
2 c_u \cot{\theta}\right) \Delta u
\label{Df}
\end{equation}

Let us now consider the particular case of a quadrupole radiation
in the linear approximation. If the quadrupole moment of the source
is $Q(u)$, then it can be shown (see eqs. (86)--(91) in \cite{Boal})
that in the linear approximation
\begin{equation}
c = \frac{1}{2} \; Q_{uu} \; \sin^2{\theta}
\label{lcu}
\end{equation}
and
\begin{equation}
- m_u = \frac{2}{15} \; Q_{uuu}^2
\label{mu}
\end{equation}
Thus
\begin{equation}
\Omega = - \frac{1}{2r} \; \sin{2\theta} \; Q_{uuu} + O(r^{-n})
\label{OME}
\end{equation}
or
\begin{equation}
\Omega = \sqrt{\frac{15}{8}} \; \frac{\sin{2\theta}}{r} \;
\left(-m_u\right)^{1/2} + O(r^{-n})
\label{Ome}
\end{equation}
linking directly the rate of precession to the rate of loss of mass.

\section{The gyroscopic preccession of order ${\displaystyle \frac 1{r^2}}$}

Let us now consider the next order (${\displaystyle \frac 1{r^2}}$). We easily
obtain:
\begin{eqnarray}
\Omega & = &-\frac{1}{2r} ( c_{u \theta}+2 c_u \cot \theta) \nonumber \\
&  & +\frac 1{r^2} \left[ M_{\theta}-M (c_{u \theta}+2 c_u \cot \theta)-c c_{u
\theta}+6 c c_u \cot \theta+2 c_u c_{\theta} \right]
\label{Om2}
\end{eqnarray}

Observe that the order ${\displaystyle \frac1{r^2}}$  contains, beside the
terms involving $c_u$, a term not involving news
($M_{\theta}$). Let us now assume that initially (before some $u = u_0 =
$constant) the system is static, in which case
\begin{equation}
c_u = 0
\label{static}
\end{equation}
which implies , because of (\ref{N})
\begin{equation}
M_{\theta} = 0
\label{staticII}
\end{equation}

and $\Omega = 0$ (actually, in this case $\Omega=0$ at any order) as expected
for  a static field ( for the electrovacuum case however, this may change
\cite{bonleter}). Then let us suppose that at $u = u_0$ the system starts to
radiate ($c_u \neq 0$) until $u = u_f$, when the news vanish again. For $u>u_f$
the system is not radiating although (in general) $M_{\theta} \neq 0$ implying
(see for example (\ref{N})) time dependence of metric functions
(non-radiative motions \cite{Boal}).

In the interval $u \in$ ($u_0$,$u_f$) the leading term of the rate of
precession of the gyroscope is given by (\ref{Om}).

For $u>u_f$ there is a precession term of order ${\displaystyle \frac1{r^2}}$
describing the effect of the tail of the wave on the gyroscope. This in turn
provides ``observational'' evidence for the violation of the Huygens's
principle, a problem largely discussed in the literature (see for example
\cite{Boal},\cite{BoNP},
\cite{tail} and references therein).

Putting aside the actual technical difficulties in performing such an
experiment, it should be clear that the monitoring of the gyroscope in the
interval ($u < u_0$, $u > u_f$) should, in principle, bring out, in a clear-cut
way, the presence of gravitational radiation.

Finally, let us consider the particular case of a quadrupole radiation in the
linear approximation. We obtain for this case (note a misprint in Eq.(87) in
\cite{Boal}).
\begin{equation}
\Omega = \sqrt {\frac {15}{8}} \frac {\sin 2 \theta}{r} (-m_u)^{1/2}+\frac
1{r^2} (-3 Q_{uu} \sin 2 \theta)
\label{linear}
\end{equation}

Therefore, for $u >u_f$, the rate of precession is controled by the second time
derivative of the quadrupole moment ($Q_{uu}$).

\section{Scenarios of radiation and estimations}

Let us now present some rough estimates for $\Omega$  or $\Delta\phi$, in
different scenarios.Before doing that, some remarks are
in order.

Since our intention here is just to provide orders of magnitude of the
expected effect, we shall restrain ourselves
to the quadrupole radiation case. Also, although Bondi approach implies
axial symmetry, it is not clear that such symmetry
is present in all examples below. This may be particularly true for the
case of collisions and bremsstrahlung. In the other cases,
specially in those of gravitational collapse and supernovae, axial symmetry
is not a too stringent condition.
In the same line of arguments, it should be mentioned that in some
examples, particularly in binary systems, the signal may last
for a too long duration. Since the convergence of the series requires (see
\cite{Bo} for details)
\begin{equation}
u<2r
\end{equation}

then, in those examples, the source should be very far from the gyroscope,
in order
to assure the convergence of the series expansion. At any rate we should insist on the point that the purpose of this section
is not to propose specific scenarios for eventual experiments, but just to
provide , however vague, orders of magnitude
of the mentioned effect.

Now, in the case of quadrupole radiation (in the linear approximation)
the rate of  precession can be related to the rate of loss of
mass of the source through (\ref{Ome}).Then for the change of  orientation
of the gyroscope $\Delta \phi$, during an interval
 of time $\Delta u$ we obtain using (\ref{Dfi}),
\begin{equation}
\Delta\phi =   \sqrt{\frac{15}{8}} \; \frac{\sin{2\theta}}{r} \;
\left(-m_u\right)^{1/2} \Delta u
\label{ang}
\end{equation}

This last equation will be used to obtain estimations of $\Delta\phi$,
whenever the rate of loss of mass
(or the total radiated mass) and the time interval of radiation are
available.In some examples, when there is not a characteristic
time scale of the emission we give an estimate of $\Omega$, from
(\ref{Ome}).For simplicity the numerical factor and
the trigonometric term, in (\ref{ang}) and (\ref{Ome}), are put equal to one.

In the last three decades, a great deal of work has been done in
identifying possible sources of
gravitational radiation (see \cite{thorne} , \cite
{sources} and references therein). Here we shall present a selection of
some of them, keeping in mind all reserves mentioned above:

\begin{enumerate}
\item Bynary systems.
\begin{itemize}
\item For the binary system of two neutron stars proposed by Clark
\cite{sources}, at  10 Kpc,
we obtain $\Delta\phi \approx 2.2 \times 10^{-15}$. This quantity is
obtained for an emission  time of $1$ second, of
a binay system emitting  mass at a rate $m_u \sim 10^{47} Joules/s$.
\item  For the pulsar 1937+214 (see  \cite{barker}), one obtains  $\Omega
\approx 1.23 \times 10^{-23}$rad.s$^{-1}$. The estimated value used for its
angular velocity $4033.8 rad/s$ leads to a power
 of emission around $2 \times 10^{29} watts$, taking into account the
Landau's formula and considering a distance of $2.5 Kpc$ for
the source. Identifying the power emitted by the system with the loss of
mass $m_u$, we obtain the given above value of $\Omega$ .
\item Similar estimations may be done for the pulsar 1913+16 (see
\cite{taylor}), knowing that the power of this pulsar
is $6.4 \times 10^{23} watt$.  The estimates yield, for a distance of $5 Kpc$,
$\Omega \approx 1.13 \times 10^{-26}$rad.s$^{-1}$.
\end{itemize}
\item Gravitational collapse and supernovae
\begin{itemize}
\item According to Shapiro \cite{sources}, during the first bounce and
rebound, the efficiency of gravitational
radiation never exceeds $10^{-3}M_{\odot}$,then for an event at 10Kpc with a
duration of the order of 1ms (free fall time),
 the maximum obtained $\Delta\phi$ is of the order of $3 \times 10^{-18}$
\item The model of collapse proposed by Wilson \cite{sources}, assumes a
radiating energy (in the form of gravitational
 radiation) of the order of $10^{-2}M_{\odot}$, during an interval of time of
$100M_{\odot}$ (in relativistic units). Then for  one solar
mass , at a distance of 10Kpc, one obtains $\Delta\phi\approx 6.4 \times
10^{-18}$.
\item In the strongest massive star collapse proposed by Ostriker
\cite{sources}, the emission is of
the order of $10^{-1}$ solar masses during $10^{-1}$s, at 1 Kpc.The
resulting $\Delta\phi$ is of the
order of $2.9 \times 10^{-15}$.
\item For a model of supernovae  proposed by Braginsky and Rudenko
\cite{braginsky},
(an emission of $10^{48}$J/s, during 1 ms at 15Mpc) one obtains $\Delta\phi
\approx 4.2 \times 10^{-21}$.
\item The characteristic parameters of a stellar collapse model proposed by
Rees et al \cite{rees}, are :
 an emission of $10^{50}$J/s during $5 \times 10^{-4}$s.  In such a
collapse there is a continuous frequency espectrum up to
$\displaystyle{\frac{1}{\tau_{\rho}}}$, being $\tau_{\rho} \sim \sqrt{\pi
G} \rho$ the characteristic time of the collapse for a final state density
$\rho$. The variation of energy $\Delta E$ goes like $\Delta E \sim
\displaystyle{\frac{1}{30 \pi} \frac{Q^2}{\tau_{\rho}^5}}$ where $Q$ is the
quadrupole moment along the axis. For an event at 5
Kpc involving a star of $6 M_{\odot}$, the resulting $\Delta\phi$ is of the
order of
$7 \times 10^{-17}$.
\item Finally, let us  evaluate the emission of gravitational radiation in
a supernovae event, leading to a neutron star,
by estimating the total change of the gravitational quadrupole moment of
the source. Recent estimations of the quadrupole
 moment of neutron stars \cite{poisson}, point to values of Q of the order
of $1.03 \times 10^{37}$kg.$m^{2}$.
 If we assume that the presupernovae massive star has a quadrupole moment
of the order of the sun
( $3.85 \times 10^{42}$kg.$m^{2}$) \cite{sol}, then the total change of
quadrupole moment is $\Delta Q \approx 10^{42} Kg m^2$.
 Taking into account (\ref{mu}),(\ref{ang}), then the obtained
$\Delta\phi$, for an event at distance of 1 Mpc, during
$10^{7} years$ (Kelvin-Helmholtz time scale for a star like the sun),
 is of the order of $2.8 \times 10^{-28}$.
\end{itemize}
\item Collisions and Bremsstrahlung.
\begin{itemize}
\item The emission, during collision of two neutron stars, given by Wilson
\cite{sources} is of the order
 of $10^{-3} M_{\odot}$ during an interval time of the order of $50
M_{\odot}$.For an event
at 10Kpc we obtain
$\Delta\phi \approx 1.14 \times 10^{-18}$.
\item In the case of two black holes collision, Detweiler \cite{sources}
proposes an emission of $10^{-3} M_{\odot}$,
with a collision time of the order of $10$M. For an event at 10Kpc this
leads to $\Delta\phi \approx 6.4 \times 10^{-25}$.
\item For gravitational bremsstrahlung within the galaxy, Ostriker
\cite{sources} suggest
 an emission of $10^{-2}$ solar masses during one second, this yields a
$\Delta\phi$ of the order of $3 \times 10^{-16}$.
\end{itemize}
\item Other sources.
\begin{itemize}
\item  Following a speculative discussion about the evolution of galactic
nucleus, Ostriker \cite{sources}
 considers the possibility of destruction of $10^{8}$ neutron stars at the
center, in aproximately $10^{6.5}$ years,
emitting $10^{-2}$ solar masses in the form of gravitational radiation, an
event of this kind, at 1Mpc of distance
 would produce a total precession of the order of $3 \times 10^{-11}$.
\end{itemize}

\end{enumerate}

\section{Conclusions}
We have seen so far that a gyroscope at rest in a Bondi frame will
precess (up to order $1/r$) as long as the system radiates, the
rate of precession being given by (\ref{Om}).

Once the radiation stops (vanishing  news) the gyroscope will continue to
precess with a rate of rotation
given by the second term of (\ref{Om2}) with $c_u = 0$ .

As can be seen from estimations done above, excluding the last example, and
the example based on
the evaluation of the change in the quadrupole moment,
the most realistic scenarios point to $\Delta\phi$ `s ranging between
$10^{-15}$ and $10^{-19}$. We ignore how far are we , with the present
technology, to the accuracy required for that kind
of measurement. Nevertheless, we want to stress that it has been our main
purpose here, just to bring out such effects in the
context of Bondi formalism.

We would like to conclude with the following comment: observe that all
along our discussion we have not made
reference to specific bandwiths. This is so because all quantities in the
Bondi approach, are not defined with respect to
any specific frecuency. In this sense eq.(\ref{muI}) (and all related
quantities), have to be considered as
integrated over all frecuencies.

\section*{Acknowledgment}
It is a pleasure to thank Prof. Bondi for interesting comments.

\end{document}